# Ultra-bright single photon source based on an atomically thin material

J.C. Drawer[1,*], V.N. Mitryakhin[1,*], H. Shan[1,*], S. Stephan[1,2], M. Gittinger[1], L. Lackner[1], B. Han[1], G. Leibeling[3] F. Eilenberger[3], R. Banerjee[4], S. Tongay[4], K. Watanabe[5], T. Taniguchi[6], C. Lienau[1], M. Silies[2], C. Anton-Solanas[7], M. Esmann[1], and C. Schneider[1]

[1]Institute of Physics, Carl von Ossietzky University, Oldenburg, 26129 Germany

[2]Hochschule Emden/Leer, Emden, Germany

[3]Institute of Applied Physics, Abbe Center of Photonics, Friedrich Schiller University Jena, Jena, Germany
Fraunhofer-Institute for Applied Optics and Precision Engineering IOF, Jena, Germany
Max-Planck-School of Photonics, Jena, Germany

[4]Materials Science and Engineering, School for Engineering of Matter, Transport, and Energy, Arizona State University, Tempe, 85287, Arizona, USA

[5]Research Center for Functional Materials, National Institute for Materials Science, 1-1 Namiki, Tsukuba 305-0044, Japan

[6]International Center for Materials Nanoarchitectonics, National Institute for Materials Science, 1-1 Namiki, Tsukuba 305-0044, Japan

[7]Depto. de Física de Materiales, Instituto Nicolás Cabrera, Instituto de Física de la Materia Condensada, Universidad Autónoma de Madrid, 28049 Madrid, Spain

*The authors contributed equally to this work.

## Abstract

Solid-state single photon sources are central building blocks in quantum communication networks and on-chip quantum information processing. Atomically thin crystals were established as possible candidates to emit non-classical states of light, however, the performance of monolayer-based single photon sources has so far been lacking behind state-of-the-art devices based on volume crystals. Here, we implement a single photon source based on an atomically thin sheet of $WSe_2$ coupled to a spectrally tunable optical cavity. It is characterized by a high single photon purity with a $g^{(2)}(0)$ value as low as 4.7 ± 0.7 % and a record-high first lens brightness of linearly polarized photons as large as 65 ± 4 %. Interestingly, the high performance of our devices allows us to observe genuine quantum interference phenomena in a Hong-Ou-Mandel experiment. Our results demonstrate that open cavities and two-dimensional materials constitute an excellent platform for ultra-bright quantum light sources: the unique properties of such two-dimensional materials and the versatility of open cavities open an inspiring avenue for novel quantum optoelectronic devices.

## Introduction

Solid-state single-photon emitters are the most efficient platform to deliver close-to-ideal single photons, excelling in both high generation rate and quantum coherence. These are the fundamental properties to enable scalable quantum optical applications. In recent years, a large palette of solid-state single-photon emitters has emerged, featuring different degrees of material-processing versatility, a wide range of emission wavelengths, operation temperatures, and polarization properties[1]. Among them, atomically thin semiconductors are promising candidates for optoelectronic and quantum applications[2,3]: They combine low-cost synthesis with maximum compatibility in terms of integration into heterostructures. An example of this class of ultimately-thin materials is the inorganic transition metal dichalcogenide (TMDC) $WSe_2$, which features single photon emission from tightly localized excitons in monolayers at cryogenic temperatures[4–8]. TMDC single-photon emitters provide high quantum efficiency[9], charge tunability[6] and polarization control[10,11] and most notably, they can be seeded at precise locations by engineering local mechanical strain in the monolayer[12,13].

The success of solid-state single photon emitters relies on photonic cavities to shape the optical density of states around the emitter, i.e. increasing the spontaneous emission rate via the Purcell effect into a specific photonic mode, ensuring optimal light collection[14]. The scalable and deterministic integration of solid-state quantum emitters in photonic microcavities still is one of the most delicate tasks in quantum engineering. While techniques based on combining nano-lithography, nano-imaging and emitter site-control were widely explored[15–17], especially in the field of III/V semiconductor quantum dots, more powerful and versatile approaches were recently developed. Among those, the concept of open photonic cavities represents an *ad hoc*, fully deterministic approach for interfacing a microcavity with single photon emitters in two-dimensional materials. In these reconfigurable Fabry Perot resonators, the two opposing mirrors allow relative displacements in three dimensions of space, facilitating precise control of quantum light emission[18,19].

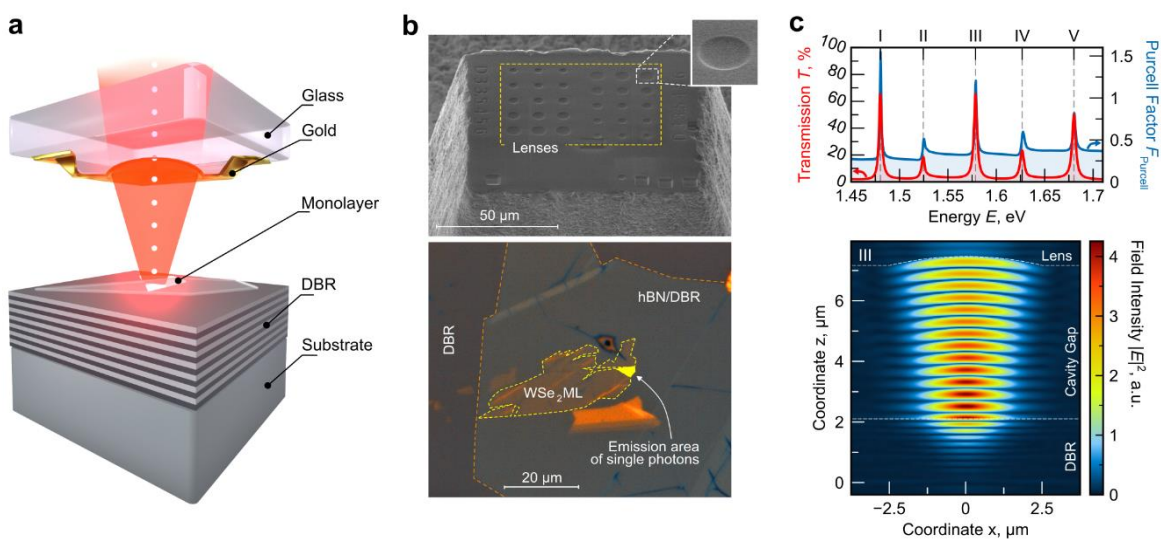

**Fig. 1 WSe$_2$ monolayer in an open cavity. a** Graphical representation of single photon emission from a monolayer source in a plano-convex open cavity under optical excitation. The relative position of the top and bottom mirror is adjustable by nano-positioners. **b** (top) Scanning electron microscope image of the mesa-type cavity top mirror with hemispherical indentations of different diameters etched by Focused Ion Beam lithography (before gold layer deposition); (bottom) optical microscope image of the WSe$_2$ monolayer placed on a SiO$_2$/TiO$_2$ DBR. The single photon source is located at the edge of the flake near a wrinkle. **c** (top) Transmission and Purcell factor (red and blue lines) of the open cavity system used in the experiments derived using FDTD simulation of the electric field of a dipole located at the monolayer position. (bottom) Real space intensity distribution inside of the cavity. The surface of the top and bottom mirror is indicated by dashed white lines.

In this work, we investigate the single-photon emission from a WSe$_2$ monolayer and deterministically couple its emission to the optical resonance of an open cavity. The reconfigurable open cavity allows us to deterministically position the single emitter of a wrinkled monolayer at the cavity center and tune the cavity resonance to the corresponding photon emission wavelength. This on-demand reconfiguration is hardly accessible in monolithic cavities such as photonic crystal cavities[20], micropillars[21], and other types of resonators[22].

The design of the open cavity sample is graphically sketched in Fig. 1a. It is based on an asymmetrical mirror design to enhance the single photon collection in the same direction as the excitation: the bottom part of the cavity consists of a distributed Bragg reflector (DBR) with high reflectivity hosting the monolayer flake, which is capped by a thin layer of hexagonal boron nitride. The top part of the cavity is built from a glass mesa containing concave hemispherical indentations of different diameters; a 33 nm thick layer of gold is evaporated onto this structure to finalize the top mirror (for further details on sample preparation see Supplementary Section 1). A scanning electron microscope (SEM) image of the cavity top mirror (before gold coating) is shown in Fig. 1b (top). The hexagonal boron nitride-capped WSe$_2$ monolayer flake on the DBR is shown in Fig. 1b (bottom). We placed the open cavity device inside a low vibration closed-cycle exchange-gas cryostat and kept it at 3.2 K.

To assess the possible performance of our cavity device, we performed Finite-Difference Time-Domain (FDTD) simulations of the experimental resonator configuration. Figure 1c (bottom) shows the resonant real space intensity distribution inside the open cavity at a wavelength of 786 nm. For a lens diameter of 5 µm and 300 nm depth (corresponding to a radius of curvature of 6.8 µm), the field is laterally confined to a diameter of ~2 µm at the emitter position. Figure 1c (top) shows the calculated transmission through the top mirror for a point dipole source and the corresponding Purcell factor as a function of wavelength.

Interestingly, the simulation predicts an on-resonant Purcell enhancement of up to 1.5 (blue line in Fig. 1c), in conjunction with an off-resonant suppression of spontaneous emission up to a factor of 3.8. The latter is a clear indicator of the strong suppression of so–called leaky modes in our cavity implementation. From our simulation, we can directly anticipate photon extraction efficiencies (also referred to as "first lens brightness") beyond 65 %.

## Results

The experimentally studied quantum dot (QD) like emitter, which evolves in our WSe$_2$ monolayer, emerges at an emission energy of 1.5707 eV (789.3 nm). It is interesting to note, that this wavelength is very close to the technologically relevant Rb-87-D2 line, with the potential for a quantum memory in future repeater networks[23]. The spectral linewidth of the QD is limited by the resolution of our detection system of 200 µeV (see supplementary Fig. S4 for a high-resolution spectrum). As a first important parameter of our source, a polarization-resolved measurement, carried out without the top mirror of the cavity, (Fig. 2a) reveals that our QD emitter displays close to perfect linear polarization up to a degree of 98.4 ± 1.3 %. We attribute this remarkable feature to the emergence of the studied QD from a monolayer wrinkle[10], creating a local and quasi one-dimensional strain-potential[24,25] which results in a strongly aligned dipole. As a next step, we add the top mirror and study the performance of the coupled cavity-emitter system. We first use non-resonant continuous wave laser excitation (532 nm) and record the sample PL for a continuously varying cavity length. The resulting color heat map is plotted in Fig. 2b. In our experiment, we observe longitudinal mode families, each consisting of three transversal modes, separated by a spacing of 26.3 meV. The modes are visualized by the guide to the eye in Fig. 2b, and emphasized in a different representation in Supplementary Section 3.

The quality factor of these cavity resonances is around 600 for the chosen mirror separations of ~5.5 µm (the separation is extracted from the longitudinal mode spacing). Importantly, and as reflected in the simulation in Fig. 1c, the mode of lowest transverse order is a Gaussian mode, which is optimally suited for coupling to a commercial single-mode fiber.

The open nature of the cavity allows us, in a straightforward manner, to study the coupled cavity-emitter system under various detunings, by changing the resonance condition via the cavity resonator length. As it is reflected in Fig. 2b, on resonance, the photoluminescence intensity of the emitter is enhanced by more than a factor of 10, which clearly reflects the strong impact of the resonator structure on the performance of the coupled emitter-cavity system. To further improve the performance of our source, we optimize the photon injection efficiency into our WSe$_2$ QD by choosing a *double resonance condition* (see Fig. S3 in Supplementary). While we maintain the spectral resonance conditions between the emitter zero-phonon line and the cavity mode, we tune our pulsed excitation laser (2 ps pulse length, 76.2 MHz repetition rate) on resonance to the next higher order longitudinal cavity mode spectrally located at 740 nm, with identical (lowest) transverse order. This condition allows us to inject light efficiently into the cavity to pump the emitter quasi-resonantly into a higher resonance shell[8].

To quantify the enhancement of spontaneous emission in our device more rigorously, we perform time-resolved photoluminescence measurements under varying emitter-cavity detunings. For these experiments, the QD emission line has been optically filtered via a coarse bandpass (~2.5 meV bandwidth) and was directly detected by an avalanche photodiode connected to a time-correlator. The corresponding decay dynamics for the off- and on-resonant case are shown in Fig. 2c (left). The characteristic decay times have been fitted with a single exponential decay function and unambiguously reflect the speed up of the spontaneous emission rate in the resonant case. The overall detuning dependence of the spontaneous emission decay is plotted in Fig. 2c (right), reflecting the interplay of the spontaneous emission rate with the optical resonance bandwidth. In addition, we have also analysed the lifetime of our emitter without the top mirror, yielding a decay time of 2.3 ns. The values in Fig. 2c, therefore, indicate a cavity-induced reduction in lifetime of 25 % on resonance, whereas off resonance the emitter experiences a more than twofold inhibition of spontaneous emission due to the presence of the open cavity. This observation is in excellent agreement with the theoretically predicted changes in lifetime shown in Fig. 1c. The predicted ratio of enhanced over-inhibited emission is 3, whereas in the experiment we find 2.71 ± 0.08.

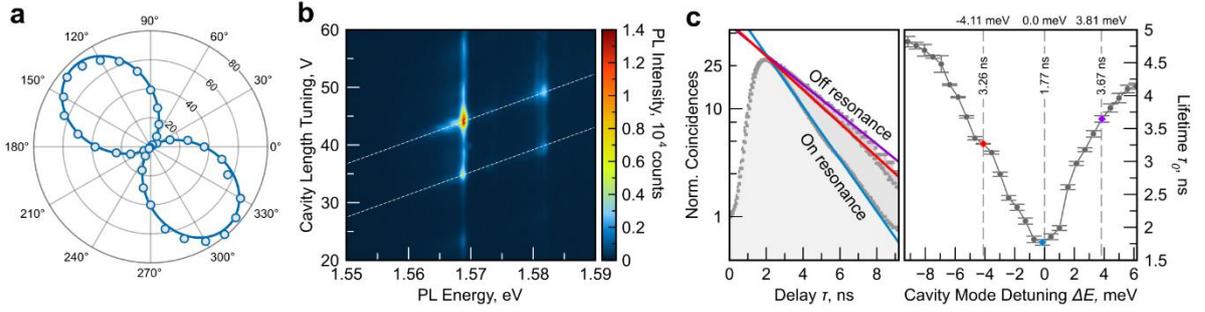

**Fig. 2 Cavity mode detuning dependent PL and lifetime. a** Polar plot of polarization-resolved PL intensity of the emission under 532 nm continuous wave excitation. The sinusoidal fit reveals a degree of linear polarization of 98.4 ± 1.3 %. **b** Colormap of PL spectra over the cavity optical length tuning while the sample is strongly excited above bandgap and outside of the stopband of the microcavity by a 532 nm continuous wave laser. Cavity modes are highlighted by dashed lines. **c** (left) Lifetime in on- and off-resonant cases (blue: on resonance; red: -4.11 meV detuning; purple: 3.81 meV detuning); (right) cavity mode detuning dependent radiative lifetime. Error bars in the right panel represent the standard error resulting from fitting the time-resolved photoluminescence data (as shown in the panel to the left) with an exponential decay function. The fitting method utilizes a damped least squares algorithm.

To assess the purity of the single photon pulses emitted by our device, we measure the second order correlation function via a standard Hanbury-Brown-Twiss (HBT) setting. From the correlation histogram in Fig. 3a, we can extract a photon anti-bunching of $g^{(2)}$ = 4.7 ± 0.7 % (details on the analysis can be found in supplementary). It is worth noting, that in these experiments the emission was only filtered by a course bandpass (~2.5 meV bandwidth); this further emphasizes the emission purity of the open cavity device.

A critical parameter in the performance of single photon sources is the probability to deliver a single photon state per excitation pulse, which is usually benchmarked by the brightness at the first collection lens. To quantify this critical performance indicator, which is of central importance for quantum communication implementations, we study the emission flux of the single photon sources as a function of the pulsed pump power (76.2 MHz repetition rate). As shown in Fig. 3b, we detect more than 1 MHz of single photon counts in our single photon detectors. After carefully assessing the transmission and detection efficiencies of our collection setup (see Table 1 of supplementary), this value directly translates into a record first lens brightness of 65 ± 4 % in a linearly polarized mode. It is worth noting that this value approaches the current state of the art in solid-state single photon sources based on III/V QDs[26–28], and widely outperforms monolayer-based triggered single photon sources reported in any implementation[29–31].

The final benchmark of the quantum-optical properties of our TMDC single photon source is the temporal coherence of the emitted single photon wavepackets, reflected by their capability to display genuine quantum interference. This property is of paramount importance for quantum applications since it guarantees the capability of photons to interfere, and thus, propagate entanglement along quantum nodes. It is furthermore of fundamental interest, since such quantum interference from atomically thin emitters has not been observed thus far.

We implement the two-photon Hong-Ou-Mandel (HOM) interference in a path-unbalanced Mach-Zehnder interferometer (see the scheme of the setup in Fig. S1 of supplementary), interfering two photon wavepackets successively emitted by the source (with an initial temporal delay of 13 ns, and eventually corrected by the delay in the interferometer). The quantum interference is extracted via the measurement of the second order correlation function between the two detectors at the output of the interferometer ($g^{(2)}_{HOM}$). The perfect bosonic quantum interference features a complete antibunching ($g^{(2)}_{HOM} = 0$).

To quantify the quantum interference from the photons emitted by our source, we measure the HOM correlation between photons with parallel/orthogonal polarizations ($g^{(2)}_{HOM,HH}$ / $g^{(2)}_{HOM,HV}$). Figure 3c shows the corresponding normalized correlation histograms, measured using the identical excitation conditions as in the HBT measurement.

We compare the critical cases of photons of orthogonal polarization (HV) in the two interferometer arms) versus parallel polarization (HH). As we reduce the width of the temporal selection window from 3 ns down to 1.1 ns – approaching the resolution limit of our detection setup – a significant difference between the parallel/orthogonal polarization correlations cases of the $g^{(2)}_{HOM}$ measurement arises consistently. Such an effect is depicted in the panel insets, where $g^{(2)}_{\text{HOM,HH}}$ and $g^{(2)}_{\text{HOM,HV}}$ correlations display different values beyond the standard deviation of the correlation peaks. This fact is further visualized in Fig. 3d, where we depict the interference visibility $V = (g^{(2)}_{\text{HOM,HV}} - g^{(2)}_{\text{HOM,HH}})/g^{(2)}_{\text{HOM,HV}}$ as a function of the post-selected temporal window.

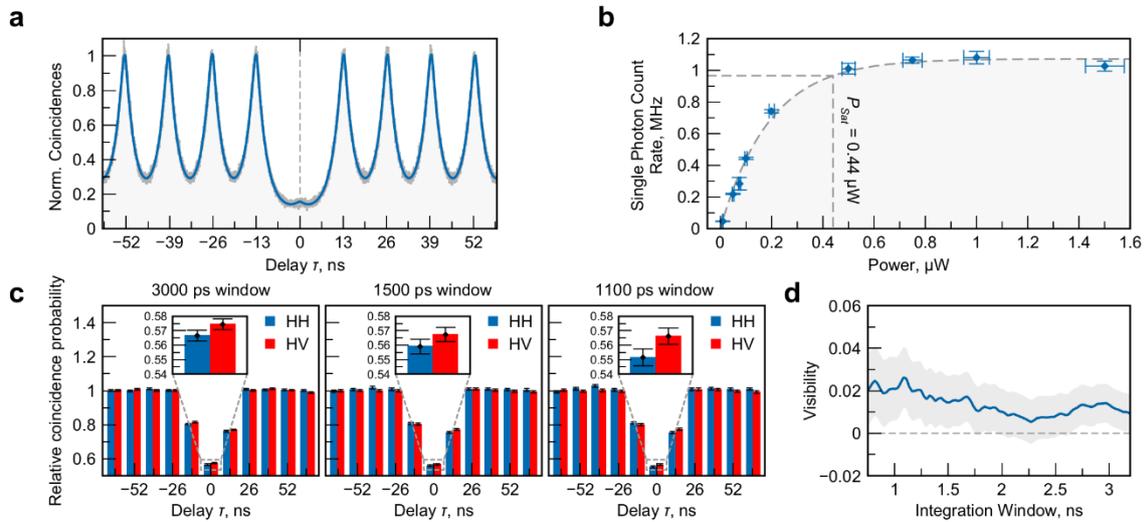

**Fig. 3 Single photon source characterization. a** Second order autocorrelation function of single photons measured in an HBT experiment with 76.2 MHz pulsed excitation in the saturation regime. The data is fitted by a double exponential decay convoluted with the system response function (details in supplementary). **b** Brightness of the source as a function of optical pump power measured before focusing onto the sample. Errors are the standard errors as the result of averaging over 10 samples for each point. **c** Second order correlation function of single photons in an HOM setup discretized by the pulsed excitation for three different temporal post-selection window sizes in case of parallel (HH) and perpendicular (HV) polarization. Error bars show standard errors resulting from fitting. **d** HOM interference visibility for varying temporal post-selection window size; the shaded area shows the confidence interval calculated from the assumption of a Poissonian distribution of counts in the integrated windows.

## Discussion and Conclusion

The results on photon quantum interference manifest the presence of a substantial dephasing channel in the TMDC QD. From the modest interference visibility, we can estimate a dephasing time of 45 ps (see Methods), which is consistent with previous studies of the linewidth of WSe2 QDs[32]. The dephasing most likely roots from rapid surface-induced charge noise and is only partly mitigated by the capping of our monolayer by hexagonal boron nitride. Thus, in order to further improve the coherence time of the QD emission, we suggest to further stabilize the charge environment via including graphene contacts to gate the system[33]. We furthermore believe, that it will be possible to boost the Purcell-enhancement in our cavity beyond a factor of 10 via minimizing the mode volume (e.g. further closing the cavity) and slightly improving the cavity quality factor (e.g. by employing a top mirror with slightly enhanced reflectivity[26] or resorting to nanoscale resonators[34].

It is further worth mentioning, that harnessing the high brightness and purity of our source, it is readily applicable in quantum communication schemes which do not rely on quantum interference and entanglement, such as the BB84 protocol in urban networks[31].

## Methods

**Sample preparation.** The sample represents itself as a Fabry-Perot type cavity consisting of two separated mirror pieces. The bottom part of the cavity is a distributed Bragg reflector made of 10 pairs of $TiO_2$/$SiO_2$ layers. The thicknesses of the layers are 85 and 131 nm respectively. This configuration corresponds to a high reflectivity region (so called stop band) centered at 755 nm. On top of this DBR, an atomically thin layer of $WSe_2$ is placed via the dry-gel stamping method. A layer of hexagonal boron nitride serving as capping layer is transferred onto the bottom part of the sample. The upper part of the open cavity consists of a $SiO_2$ substrate, into which a square mesa with dimensions of 100 μm x 100 μm is milled. A focused ion beam (FIB, FEI Helios 600i) was used to mill lenses with diameters of 3, 3.5, 4, 5 and 6 μm and identical depth of 300 nm into the mesa surface. Finally, a 33 nm layer of gold was evaporated onto this structure. Since the upper gold mirror prevents visualization of the bottom mirror, a 20 um x 30 um window was etched by FIB, removing the gold on the mesa. The top and bottom mirrors are mounted onto separate motorized stages featuring nanopositioning in XYZ directions. The whole device is located in a low-vibration closed cycle cryostat, allowing high stability of the cavity alignment at low temperatures.

**Optical Spectroscopy settings**. All measurements have been conducted in a confocal microscope setup with excitation and collection through a lens in the top part of the sample. A simplified partial sketch of the setup is shown in Supplementary Fig. S1. The collected emission is filtered by a longpass filter ensuring that the laser reflected into the path in not present in the spectrum. The emission is then guided with a set of mirrors and projected onto the slit of a monochromator, diffracted and recorded by a Peltier-cooled charge coupled device.

To isolate the single photon emission, the spectrum is accordingly filtered with a set of tunable (within the range of 710-800 nm) shortpass and longpass filters achieving bandpass windows down to 2-3 meV. In the HBT setup configuration, the emission is coupled via a zoom collimator to a single mode fiber that is then connected to a 50:50 fiber beam splitter. Its two outputs are each connected to an avalanche photodiode as a single photon detector. A time-correlation device receives the signals from both APDs and the coincidences of the events within a certain bin size are recorded. The setup configuration for the study of Hong-Ou-Mandel type interferences is detailed in Supplementary Section 1.

**Modelling**. The electromagnetic field inside the cavity has been simulated via placing a dipole emitter on the bottom mirror of the cavity, utilizing the commercial software Lumerical. Monitoring the power radiated from the dipole throughout the structure allowed us to derive the Purcell factor, the transmission coefficient of the upper part of the cavity and to acquire the field distribution for specific wavelengths. A description containing all the technical details of the simulations is located in Supplementary Section 2.

The dephasing time is estimated from the interference contrast via V~T2/(2*T1). We notice that without correcting for the finite $g^{(2)}$ value in the HBT experiment, this yields a conservative estimate of T2. To account for the finite integration time window, the relaxation rate can be replaced by the temporal width of the integration window.

# Acknowledgement

This project was funded within the QuantERA II Programme that has received funding from the European Union's Horizon 2020 research and innovation programme under Grant Agreement No 101017733, and with funding organisations the German ministry of education and research (BMBF) within the projects EQUAISE and TubLan Q.0.

Financial support from the European Research Council within the project unLimit2D (Grant number 679288) is acknowledged. Furthermore, the open cavity was developed with support of the projects SCHN1376 11.1 and SCHN1376 14.1, funded by the German Research Foundation (DFG). We also thank the DFG for support within the program for major equipment (INST 184/220-1 FUGG).

M.E. acknowledges funding by the University of Oldenburg through a Carl von Ossietzky Young Researchers' Fellowship.

S.T acknowledges primary support from NSF DMR 2111812 for materials development, NSF GOALI 2129412 for scaling, and NSF ECCS 2111812 fabrication. We acknowledge partial support for DOE-SC0020653 (materials texture development), NSF ECCS 2052527 for electronic and NSF DMR 2206987 for magnetic purity tests.

C.A.S. acknowledges the support from the Comunidad de Madrid fund "Atraccion de Talento, Mod. 1", Ref. 2020-T1/IND-19785 and the project from the Ministerio de Ciencia e Innovación PID2020-113445GB-I00.

**Supplementary information:**

# Ultra-bright single photon source based on an atomically thin material

## Supplementary section 1: Experimental setup details

**Photoluminescence measurements**. Excitation is done either by a Coherent Mira Optima 900-F mode-locked Ti:Sapphire laser or a 532 nm diode laser. All measurements are performed with the lower mirror-sample-part mounted on a motorized XYZ stage, model Attocube ANPx101/LT (for x, y) and ANPz51/RES/LT (for z). This assembly is mounted together with the upper cavity mirror on a motorized XY stage, model Attocube ANPx311/LT/HV, in a closed-cycle cryostat, model Attocube attoDRY1000, operating at 3.2 K. The first lens above the sample is a Thorlabs 354105-B with NA = 0.6 mounted on a motorized Z stage, model Attocube ANPx311/RES/HL/LT, inside the cryostat. The sample emission in photoluminescene measurements is recorded by a charge coupled device (CCD), model Andor iKon-M 934 Series, attached to a spectrometer, model Andor Shamrock SR-500i. The CCD is used at highest sensitivity (50 kHz read out rate, 4x pre-amplification, and sensor temperature -85 °C); the exposure time is set to 1 s.

**Correlation and time-resolved measurements**. The single photon emission line is spectrally filtered by the following filters:

- Semrock 790 nm VersaChrome Edge™ tunable longpass
- 790 nm VersaChrome Edge™ tunable shortpass
- Thorlabs FEL0750 longpass
- Thorlabs FESH0800 shortpass

The resulting signal is then coupled into a 5 µm single mode fiber (model Thorlabs P3-780AR-2) via a zoom collimator (model Thorlabs ZC618APC-B). For the Hanbury-Brown-Twiss (HBT) measurement, its output is coupled to a 50:50 single mode fiber beam splitter (model Thorlabs TW805R5A2). For the Hong-Ou-Mandel (HOM) measurement, both free space and fiber beam splitters are used. The sketch of the setup for HOM measurements is shown in Fig. S1.

In both cases, HBT and HOM, the outputs of the fiber beam splitter are each connect to one Laser Components COUNT® T COUNT-T100-FC avalanche photo diode, which are connected to a quTAG time tagger for correlation measurement. The timing jitter of the photo diodes is specified at 500 ps. We measure the system response function by coupling the pulsed laser into the HBT setup and use it to deconvolute the single photon correlation measurement in the main text. To obtain the photon purity, the HBT correlation histogram (see Fig. 3a) is fitted by a train of equidistant double exponential decay functions with equal decay time and individual peak height, which is convoluted with the measured system response function. A constant is added as a fit parameter to account for a noise background. To obtain the value of $g^{(2)}(0)$, the ratio of the height of the peak at zero delay and the average height of all other peaks is calculated.

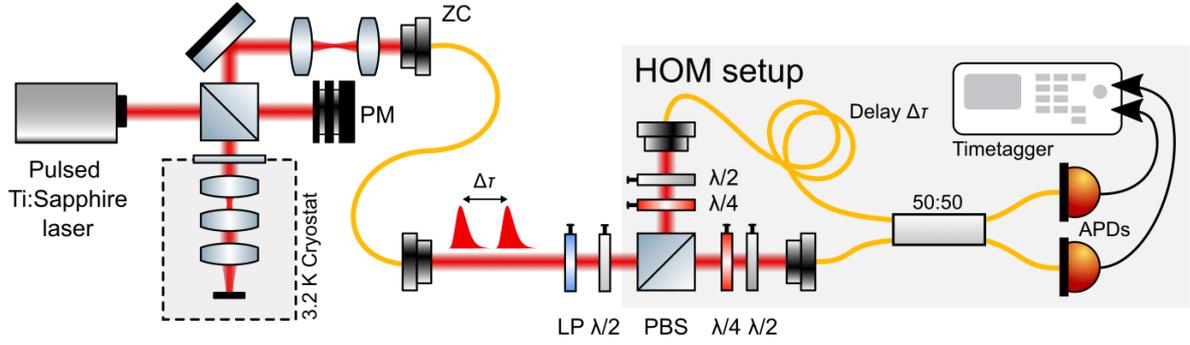

**Fig. S1** Sketch of the HOM interferometer setup. The sample is put in a closed-cycle cryostat and kept at 3.2 K temperature while being irradiated by a pulsed Ti:Sapphire (Coherent Mira Optima 900-F) laser. The emission coming from the sample is then spectrally filtered and guided through a zoom collimator (ZC) to the Mach-Zehnder interferometer. In the collection path, a pair of lenses is set as a telescope to adjust the beam size, thus improving the efficiency of fiber coupling. On the input of the interferometer a linear polarizer (LP) and a half waveplate are set in front of a polarizing beam splitter (PBS). In this way, both arms of the interferometer can be balanced in power. Then, the output beams coming from the PBS pass through fine- polarization control optics (a quarter- and half-waveplate) that account for imperfections of the PBS. This allows us to be sure that emission in both arms has a high degree of linear polarization. One of the half waveplates is used to rotate the polarization of one arm with respect to the other to achieve the parallel (HH) or orthogonal (HV) polarization. In addition, one arm is delayed by the excitation laser repetition period. PM: powermeter, ZC: zoom collimator, LP: linear polarizer, PBS: polarizing beam splitter, λ/2 (λ/4): half-waveplate (quarter-waveplate), APDs: avalanche photo diodes.

## Supplementary section 2: FDTD simulation

We perform FDTD simulations (Lumerical) of the open cavity device. The bottom mirror of the cavity is simulated as 10 alternating $SiO_2/TiO_2$ layers placed on a substrate with refractive index $n$ = 1.5 (glass), starting with $TiO_2$. The refractive indices and layer thicknesses are taken from data provided by the mirror manufacturer Laseroptik GmbH: $n_{SiO2}$ = 1.45, $d_{SiO2}$ = 131 nm, $n_{TiO2}$ = 2.28, $d_{TiO2}$ = 85 nm. The $WSe_2$ flake is simulated as a thin layer of a high refractive index dielectric with $n_{WSe2}$ = 4.3, i.e. the refractive index at the energy of the emitter. A 5 nm thick layer of hBN with $n_{hBN}$ = 2.25 is placed on top. The top mirror of the cavity consists of a spherical lens of maximum depth of 300 nm and a diameter of 5 µm, etched into a substrate with $n$ = 1.5 (glass). The lens is, as in the experiment, covered with a thin gold layer of 33 nm thickness. The dielectric function of gold at cryogenic temperatures is taken from Ref. [1].

The cavity modes are excited by a broadband dipole emitter that is placed inside the $WSe_2$ layer, with the dipole moment oriented along the *x*-axis. The dipole is surrounded by a set of monitors to measure the total power radiated by the dipole to record the energy-dependent Purcell factor. The spatial field profiles are recorded in the *xz*-plane. Radiation leaving the cavity, either through the top or bottom, is recorded by transmission monitors to compute the extraction efficiency. The transmission monitors are located 400 nm above the gold surface and 200 nm below the DBR.

The total simulation size is 7.5x7.5x8.8 µm$^3$ and the simulation runs for 3.5 ps such that the fields inside the cavity have sufficiently decayed. A spatially variable meshing is applied with cell sizes ranging from approx. 50 nm in free-space to 2 nm for the $WSe_2$ layer and 3 nm for the gold film of the top mirror. Symmetric and anti-symmetric boundary conditions are applied in order to reduce the run-time and memory demand of the simulation.

The simulations allow extracting the energy-dependent Purcell factor (i.e. the acceleration of the spontaneous emission of the emitter) and the cavity transmission (relevant for the double resonance excitation conditions that we put in place) through the 5 µm diameter lens etched in the top mirror. In the investigated spectral range, we observe the formation of Fabry-Perot modes, which are identified by peaks in both transmission and Purcell factor. For the mode labelled III in Fig. 1c of the main text, we study the spatial mode profile by analyzing the real space distribution of the electric field intensity (Fig. 1c, bottom). It shows a strong confinement of the electromagnetic field in the xy-plane to a diameter of ~2 µm due to the micro-lens structure.

## Supplementary section 3: Additional data

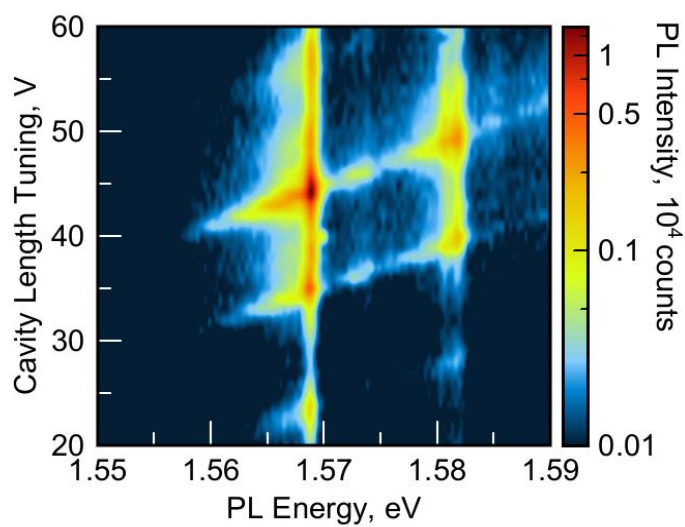

**Fig. S2** Colormap of PL spectra under excitation of 532 nm laser. The data is reproduced from Fig. 2b in the main text, which is plotted with a linear false color-scale. To clearly exhibit the cavity modes, here we encode the data with a logarithmic color-scale. The bright inclined lines represent optical modes of the open cavity device.

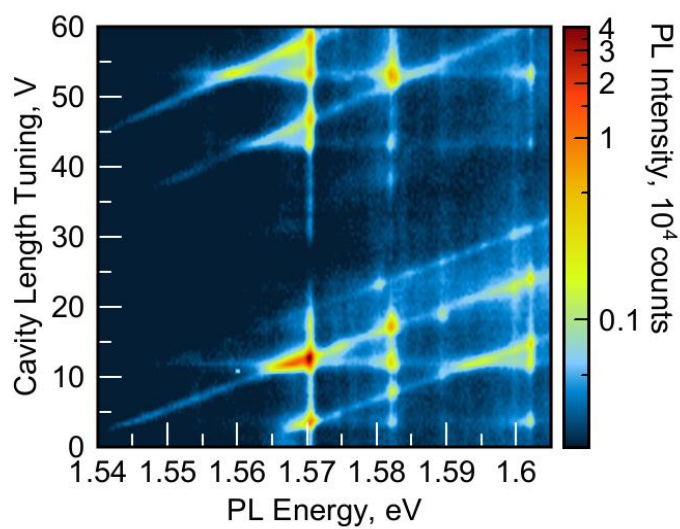

**Fig. S3** Colormap of PL spectra as a function of the cavity detuning and the emission energy, excited by 740 nm pulsed laser. In comparison to the case of 532 nm excitation (Fig. S2), the newly emerging horizontal features are observed (e.g. for cavity length corresponding to 54 and 12 V DC applied to the piezo of one of the Z direction nanopositioners). This indicates a cavity mode being in resonance with fixed-wavelength laser excitation. When the excitation laser and the PL emission simultaneously match in frequency with a cavity mode, the system is in double resonance for which the PL enhancement is most pronounced.

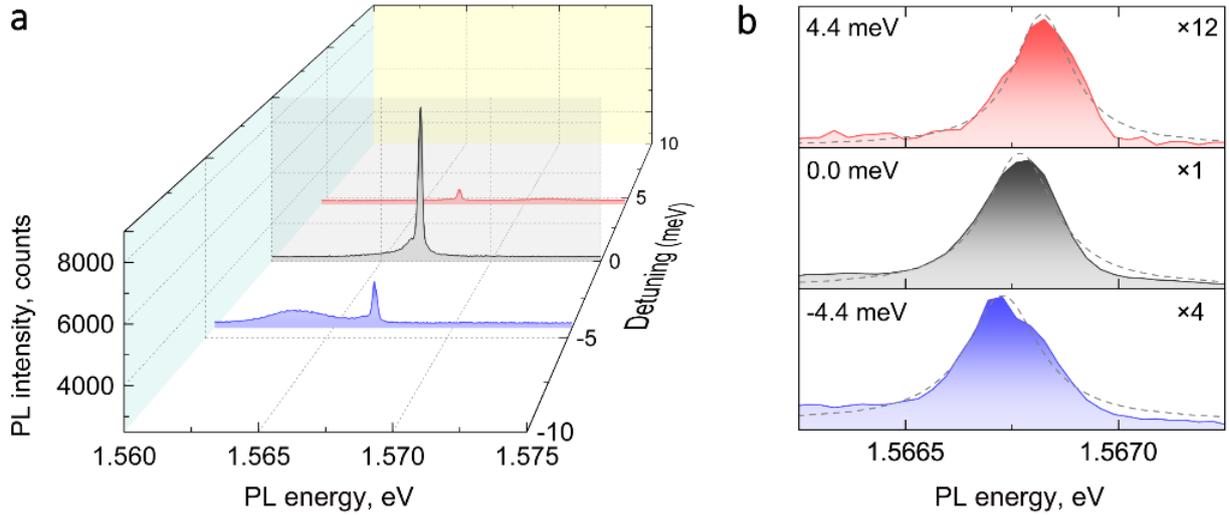

**Fig. S4 Representative spectra at different detuning values under double resonance condition. a** PL spectra of single photon emission, at resonance (gray), positive (red) and negative (blue) detuning. Laser excitation at 740 nm is resonant with one of the optical modes of the open cavity. **b** Zoomed-in spectra of the single photon emission peak area in the spectra shown in panel **a**. The peaks are fitted with a Lorentzian function (dashed lines). The resulting linewidth is 200 µeV (extracted for zero detuning).

**Experimental setup calibration.** In Tab. S1 we list all contributions to the optical losses in our experimental setup.

|  | Value | Abs. Error |
|---|---|---|
| **Cryostat (cryostat window, 3 lenses) and beamsplitter transmission (*)** | 22.87 % | 0.05 % |
| Cryostat (cryostat window, 3 lenses) transmission | 45.22 % | 0.10 % |
| Beamsplitter transmission | 50.57 % | 0.14 % |
| **Free space path transmission (2 lenses, 7 mirrors)** | 29.29 % | 0.14 % |
| **Spectral filter transmission and fiber coupling efficiency (*)** | 50.4 % | 1.9 % |
| Spectral filter transmission | 79.9 % | 2.0 % |
| Fiber coupling efficiency | 63.0 % | 2.2 % |
| **Avalanche photodiode efficiency** | 64.3 % | 2.2 % |
| **Total efficiency** | **2.17 %** | **0.11 %** |
|  |  |  |
| **Dead time corrected count rate of the APD** | 1080 kHz | 40 kHz |
| **Excitation laser repetition rate** | 76227.93 kHz | 0.18 kHz |
| **Quantum efficiency** | **65 %** | **4 %** |

(*): combined measurement to reduce the error

**Tab. S1 Calibration for losses in the optical setup.**